\documentclass[12pt]{article}

\usepackage{amsmath}
\usepackage{soul}
\usepackage{amsfonts}
\usepackage{amssymb}
\usepackage{physics}
\usepackage{color, soul}
\usepackage[left=3cm,right=3cm,top=3cm,bottom=3cm]{geometry}
\usepackage{graphicx}
\usepackage[indent=0pt]{parskip}
\usepackage{tikz}
\usepackage{array}
\usepackage{booktabs}
\usepackage{xcolor}
\usepackage{times}
\usepackage{float}
\usepackage{comment}
\UseRawInputEncoding
\usepackage[numbers,sort&compress]{natbib}

\usepackage{graphicx}    
\usepackage{subcaption}  
\usepackage[percent]{overpic}     
\usepackage{xcolor}      
\usepackage{caption}     
\usepackage{ulem}

\newcommand{\pdfrac}[2]{\frac{\partial #1}{\partial #2}}

\newcommand{\AAu}[1]{\textcolor{black}{#1}}
\newcommand{\AAub}[1]{\textcolor{black}{#1}}

\newenvironment{sciabstract}{%
\begin{quote} \bf}
{\end{quote}}

\newcounter{lastnote}

\title{Regulation of propulsion in assemblies of thermophoretic nanomotors}

\author
{Y. De Figueiredo$^{1}$, U. Delabre$^{1}$, S. Cassagn\`{e}re$^{1}$, M. Romanus$^2$,\\ J-P. Delville$^1$, M-H. Delville$^2$, A. Aubret$^{\ast1}$\\
\\
$^{1}$CNRS-LOMA, UMR 5798, 351 Cours de la Lib\'{e}ration, F-33400 Talence, France\\
$^{2}$CNRS, University of Bordeaux, Bordeaux INP, ICMCB, \\ UMR 5026, 87 Avenue du Dr. A. Schweitzer, F-33608 Pessac, France\\
\\
\normalsize{$^\ast$To whom correspondence should be addressed; E-mail: antoine.aubret@cnrs.fr}
}
\date{}

\begin{document}
\baselineskip16pt


\maketitle


\begin{sciabstract}
Active particles locally transduce energy into motion, leading to unusual and emergent behaviors. However, current synthetic particles lack sensing and adaptation mechanisms. Here, we demonstrate a novel regulation pathway, through the combined use of thermophoretic propulsion and nanometric building blocks. We build an active fluid composed of artificial nanomotors and study its three-dimensional (3D) dynamics. We use laser-induced photo-thermal effect to actuate nanoparticles, and probe their self-propulsion within assemblies. Despite significant thermal fluctuations at the nanoscale, our results reveal a strong dependence of the thermophoretic propulsion on the concentration of nanomotors, leading to ultrafast velocities of up to $\sim 800$ $\mu$m/s. This unique behavior originates from a strong coupling of the local concentration of nanomotors and the temperature field, which feeds back on the thermophoretic mobility of the nanoparticles. We rationalize our results from independent modeling of all thermal effects, accounting for nonlinearities of thermophoretic self-propulsion. Our results open novel routes for the design and self-regulation of 3D active fluids by thermal processes.
\end{sciabstract}
\newpage

\section*{Introduction}

\color{black}
Active particles locally transduce energy to propel, exhibiting behaviors significantly different from equilibrium systems \cite{Bechinger_Volpe-RevModPhys-2016}. 
At high density, in particular, active particles present emergent behaviors like dynamic phase separation \cite{Palacci_Pine-Science-2013,Theurkauff_Bocquet-PRL-2012}, active turbulence \cite{Peng_Cheng-SciAdv-2021,Wu_Dogic-Science-2017}, or flocking \cite{Chardac_Bartolo-PRX-2021,Bricard_Bartolo-Nature-2013}. Critically, synthetic systems cannot sense their environment. In contrast, biological systems adapt their dynamics spontaneously through internal regulatory pathways \cite{Crespi-TEE-2001}. This leads to their complex organization, far beyond what can currently be achieved with artificial systems \cite{Needleman_Dogic-NatureRevMat-2017}.

Current approaches to adapt activity have relied on user-induced external stimuli such as flows, light, or electrical modulation. Novel exotic phases could be reached on demand, like specific density patterns \cite{Arlt_Poon-NatureComm-2018,Frangipane_Dileonardo-eLife-2018} or the control of the assembly of active particles \cite{Aubret_Palacci-NatComm-2021,Martinet_Palacci-PRX-2025}. Spontaneous sensing and adaptation with synthetic particles remains however hardly achieved \cite{Lefranc2025,Alvarez_Isa-NatureComm-2021}, and have been mainly realized through computer-feedback mechanisms \cite{MuinosLandin-Cichos-ScienceRobotics-2021,Lavergne_Bechinger-2019,Baeuerle_Bechinger-NatComm-2018,FernandezRodriguez_Isa-NatureCommunications-2020}. Hence, designing autonomous adaptation processes in synthetic active matter is an open challenge.

While synthetic systems have mostly relied on the use of microscopic building blocks, active nanoparticles (NPs) offer novel opportunities for the design of responsive active systems. Their low mass enables 3D propulsion, with enhanced efficiency \cite{Sabass_Seifert-PRL-2010,Bregulla_Cichos-FaradayDiscuss-2015}, and velocities far exceeding those of microparticles \cite{Wang_Mallouk-ACSNano-2021,Lee_Fischer-NanoLetters-2014,Qin_Fan-AngewandteChemie-2017,Christoulaki_Buhler-PRE-2025,Truong_Grelet-Nanoscale-2026}. Besides, due to their volume distribution, they can significantly modify the bulk properties of a fluid, provided a high enough concentration can be reached. For instance, catalytic nanomotors can exhibit collective swarming as a result of self-induced solutal convection currents \cite{Choudhury_Fischer-AdvancedMaterials-2019,Chen_Sanhez-NatComm-2024, Hortelao_Sanchez-SciRob-2021}. These studies demonstrate the potential of nanomotors to significantly alter their environment, coupling their particle number density to activity. However, studies have been limited to catalytic nanomotors, without adaptation mechanisms.

\AAu{Here, we unveil new adaptive pathways to motility, through the combined use of thermophoresis and temperature-sensitive nanomotors. We synthesize Au/SiO$_2$ nano-heterodimers (size $\sim 60$ nm), and use photothermal heating to trigger their thermophoretic motion. We quantitatively analyze their dynamics at various concentrations using time-correlated spectroscopic measurements. Our results show a strong dependence of the phoretic speed on the local concentration of nanomotors, with observed ultra-fast velocities reaching $\sim 800$ $\mu$m/s. We show through quantitative modeling that the self-amplification of the velocity with the concentration originates from both large-scale heating of the sample, and the strength of microscopic thermal gradients. Our results show that high particle concentration provides higher swimming efficiency compared to dilute systems.}

\color{black}

\section{Results}\label{sec:exp}

\subsection{Synthesis and design of nanomotors}
Our strategy relies on the use of self-thermophoresis induced by photothermal heating to actuate nanomotors. Self-propulsion occurs due to the asymmetric properties of the nanoparticles, which sustain their own thermal gradients following asymmetric light absorption [Fig.1]. To this end, we synthesize nano-heterodimers (NHDs) made of an absorbing gold (Au) nanosphere, on which a silica lobe is grown [Fig.1a]. We adapt a previously reported synthetic route, which relies on a site-selective nucleation method \cite{Park_Kane-Nanoscale-2018} of tetraethylorthosilicate (TEOS) onto Au seeds, following a standard St\"{o}bber growth method \cite{Stober-JCIS-1968}. Briefly, we functionalized the surface of citrate-stabilized Au nanospheres  (size $\sim 30$ nm) with two non-compatible ligands (4-MercaptoBenzoic Acid (4MBA) and Polyacrylic acid (PAA)) [Fig.1a]. By adjusting the alcohol-water cosolvent ratio, polymer weight, and the relative concentrations of 4MBA and PAA, we controlled the directional growth of silica on Au (see Materials \& Methods). This approach results in the synthesis of NHDs with tunable eccentricity, \textit{i.e.}, core-shell nanoparticles in the absence of PAA, and net heterodimers at a specific PAA concentration [Fig.1b and 1C]. The particles present a colloidal stability in a [1:1] molar ratio of water/2-propanol mixture, and the final size of the NHDs typically ranges between $\sim 40$ and $70$ nm, for an aspect ratio between long and short axes of the order of $1.3$ [Supplementary Fig.1 \& 2]. Fig.1d shows a representative image of a sample of the NHDs used in the study, illustrating both high synthesis yield and reasonable monodispersity.

All the experiments are performed in standard glass capillaries. In the absence of light exposure, the NHDs exhibit Brownian motion. When illuminated by green light, the Au seed acts as a highly efficient light-to-heat nanotransducer, as a result of the absorption of light ($\lambda= 532$ nm) at the plasmon resonance wavelength [Fig.1b]. It results in a 3D self-propulsion by thermophoresis, at a speed $v \propto \mu^{NHD} \langle\nabla T\rangle$, where $\mu^{NHD}$ is the NHD thermophoretic mobility, and $\langle\nabla T\rangle$ the typical tangential temperature gradient at the surface of the NHD \cite{Jiang_Sano-PRL-2010}.

\begin{figure}[H]
    \centering
    \includegraphics[scale=1]{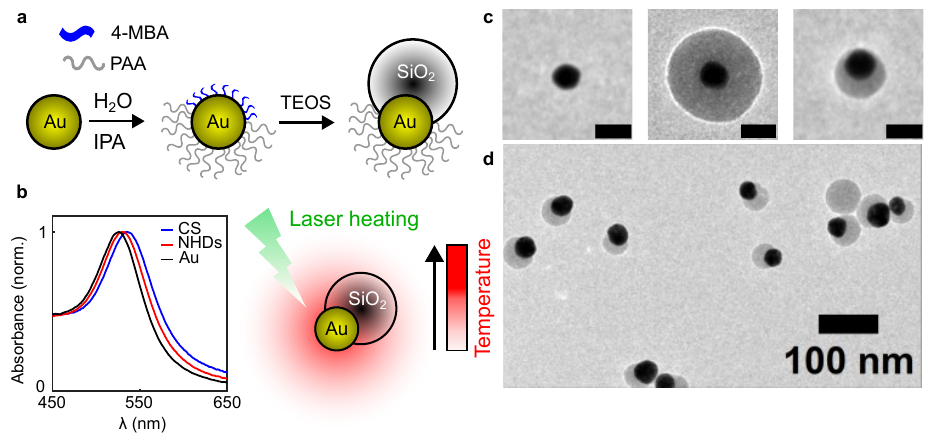}
    \caption{\textbf{Au/SiO$_2$ nano-heterodimers (NHDs) as nanomotors.} \textbf{a)} Scheme of the synthesis process. Gold nanospheres are used as a substrate for the asymmetric growth of SiO$_2$. Following the grafting of competing ligands onto the Au surface (4-MBA and PAA, see main text), the silica precursor (TEOS) nucleates a SiO$_2$ lobe on the Au sphere. \textbf{b)} \textit{Left:} Activation is performed at $\lambda=532$ nm, close to the plasmon resonance of gold, as observed on the extinction spectra of the NHDs, core-shell (CS), and Au nanospheres (Au). \textit{Right:} Experimental strategy for self-propulsion. The NHDs are activated by photothermal heating. Following laser-light absorption of the Au part, the NHDs propel in 3D by self-thermophoresis. \textbf{c)} Transmission Electron Microscopy (TEM) images of (from left to right): gold nanosphere, Au/SiO$_2$ core-shell nanoparticles (absence of PAA), and Au/SiO$_2$ NHDs (presence of 4-MBA and PAA). Scale bar is 50 nm. \textbf{d)} Large-scale TEM image of a typical sample of NHDs, showing their monodispersity.}
\end{figure}

\subsection{Experimental setup}
\subsubsection{Auto-correlation analysis}
The NHDs exhibit 3D dynamics due to their large sedimentation length $l_s$ ($l_s=k_BT/mg \sim 1.6$ mm, where $m$ is the buoyant mass of the nanoparticle, $k_B$ the Boltzmann constant, and $g$ is the gravitational acceleration). However, their sub-diffraction-limited size prevents the analysis of their motion with standard optical microscopy techniques. We thus take advantage of the known long-time diffusive dynamics of active particles to quantify their self-propulsive velocity \cite{Howse_Golestanian_PRL-2007,Palacci_Bocquet-PRL-2010_1,Lee_Fischer-NanoLetters-2014}. At times greater than the Brownian reorientation time $\tau_R \propto R_h^3$, where $R_h=32\pm 2$ nm is the mean hydrodynamic radius of the NHDs, activated nanoparticles present effective 3D diffusion, quantified by:

\begin{equation}
    D_{\text{eff}}=D + \ \frac{v^2\tau_R}{6},
    \label{eq:Deff}
\end{equation}
where $v$ is the thermophoretic self-propulsion velocity, $D=k_BT/6\pi \eta R_h $ the equilibrium diffusion coefficient, with $\eta(T)$ the temperature-dependent viscosity of the solvent, and $T$ the temperature. Hence, the measurement of $D_{\text{eff}}$ gives direct access to the thermophoretic velocity.

We use spatially resolved, confocal scattering correlation spectroscopy (CSCS) to quantify the dynamics of the NHDs, allowing us to probe small volumes ($\sim 11\pm4$ $\mu$m$^3$) and spatial variations of diffusion coefficients for different concentration regimes, going beyond previous studies \cite{Qin_Fan-AngewandteChemie-2017,Christoulaki_Buhler-PRE-2025,Liu_Ren-JACS-2014,Truong_Grelet-Nanoscale-2026}. 
The technique relies on the analysis of the light scattered by the transient crossing of the confocal volume by freely moving particles [Fig.2a] (see [Supplementary Fig.3 \& 4]). Statistical analysis of the series of scattered photon events allows us to extract $D_{\text{eff}}$: from the intensity $I(t)$ at time $t$ recorded on the photodetector, we then compute the auto-correlation function $g^{(2)}(\tau)=\langle I(t+\tau)I(t)\rangle/\langle I(t)\rangle^2$. The fluctuation of the signal coming from nanoparticle dynamics results in interfering coherent waves, giving an exponential decay to $g^{(2)}$ with a characteristic time $\tau_C=1/2q^2D_{eff}$, where $q$ is the scattering wavevector. An additional incoherent contribution comes from fluctuations in the nanoparticle number in the confocal volume, with a timescale $\tau_D=\omega^2/4D_{eff}$, where $\omega=550 \pm 50$ nm is the width of the confocal volume. We get (Material \& Methods):

\begin{equation}
    g^{(2)}(\tau)-1=\alpha e^{-\frac{\tau}{\tau_C}}+G_0\frac{1}{\Big(1+\frac{\tau}{\tau_D}\Big)\sqrt{1+k^{2}\frac{\tau}{\tau_D}}}.
    \label{eq:g2}
\end{equation} 

 Here, $\alpha$ and $G_0$ are amplitudes for the coherent and incoherent scattered wave contributions, and $k = \omega_Z/\omega = 19\pm3$ accounts for the ellipticity of the confocal volume along the axial direction, which we quantify through independent measurements [Fig.2a \& Supplementary Fig.5-8]. 

All the correlation curves are fitted using only two free parameters $G_0$ and $D_{eff}$ (see Methods \& [Supplementary Sec.3]). Critically, based on $\tau_C$ and $\tau_D$, CSCS allows us to span a wide range of NHDs concentrations, and to probe the dynamics of nanomotors in the bulk of the sample, far from interfaces [Supplementary Fig.9]. 

\subsubsection{Experimental setup}
The experimental configuration is shown in [Fig.2a] (see [Methods] and [Supplementary Fig.4] for full description, ). A 532 nm-laser is used to control the activity of the NHDs. The beam is split into two independent vertical beams propagating in opposing directions. This allows for the control of radiation pressure forces independently of heating. Both beams are focused at the back focal plane of a pair of identical microscope objectives (NA=0.5), forming collimated Gaussian beams on the sample with a waist of $\omega_g=40$ $\mu$m [Supplementary Fig.5]. This prevents significant optical gradient forces on the nanoparticles and ensures nearly uniform illumination over the probe volume, with $\omega_g\gg \omega$. We use a low-power red laser ($\lambda = 632.8$ nm, $P\lesssim10$ $\mu$W) to probe the motion without significant absorption of the gold nanospheres at this wavelength [Supplementary Fig.3 \& 10]. The red beam is focused through the lower objective, which also collects the back-scattered photons, giving $I(t)$. The scattered signal is then spatially filtered and collected on a photon detector, enabling the computation of $I(t)$ and $g^{(2)}(\tau)$ during post analysis. 
 
\subsection{Individual propulsion in the dilute regime}
We first investigate the effect of photothermal heating on the individual dynamics of NHDs. We focus on the low excitation regime and when inter-particle interactions are negligible ( $c\sim 6\cdot 10^{13}$ NPs/L, corresponding to $\sim$ 4 $\mu$m distance between the particles). This is confirmed by the independence of the measurements on the concentration when further diluting the sample. The typical number $N$ of NHDs in the confocal volume is such that $N \approx 1$, with a linear absorption coefficient of the sample $\alpha = 100$ m$^{-1}$. At equilibrium, \textit{i.e.}, in the absence of light excitation, we obtain a typical value for the NHDs $D^{NHD}_0 = 2.6\pm0.2$ $\mu$m$^2$/s, in excellent agreement with the theoretical value $\approx2.5$ $\mu$m$^2$/s, estimated from the Stokes-Einstein expression, and accounting for the size-dependency of scattering strength [Supplementary Sec.2] (considering room temperature $T_0=294$ K and $\eta_0=2.43$ mPa.s). Upon heating of the NHDs, a net shift of the correlation function can be observed, indicative of a shorter fluctuation timescale [Fig.2b]. Computing $g^2(\tau)-1$, and fitting from Eq.\eqref{eq:g2} allows us to quantitatively extract $D^{NHD}_{eff}/D^{NHD}_0$ at room temperature $T_0$. The diffusivity shows a monotonic increase with the incident power [Fig.2c]. To assess the specific effect of shape asymmetry on the evolution of $D_{\text{eff}}^{NHD}$, we calibrate our results against measurements on isotropic Au nanospheres. The dynamics of Au nanospheres under low laser excitation remains constant, with an independence of the diffusion coefficient $D^{Au}/D^{Au}_0$ with light intensity in the range of investigated low powers [Fig.2b,c]. This invariance highlights the negligible influence of localized heating effects on solvent properties and diffusivity, contrary to what has been previously suggested \cite{Rings_Klaus-PRL-2010,Truong_Grelet-Nanoscale-2026} (see [Supplementary Fig.11]).

\begin{figure}
    \centering
    \includegraphics[scale=1]{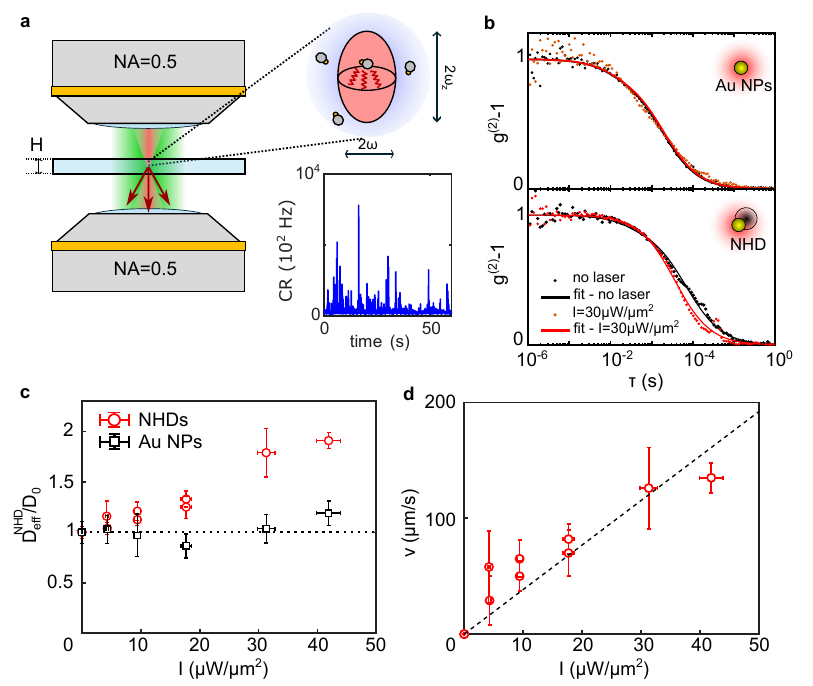}
    \caption{\textbf{Individual propulsion of nanomotors - dilute regime.} \textbf{a)} Scheme of the experimental configuration. Two green collimated laser beams excite the sample (height $H=200$ $\mu$m). The scattering from a red laser is used to probe the dynamics, with back-scattered photons collected onto photon detectors. The intensity time-trace (bottom - right panel) is reconstructed, and its auto-correlation informs about the dynamics. \textbf{b)} Representative correlation functions $g^2(\tau)-1$ for equilibrium (black dots), and excited samples (red dots) under $I=30$ $\mu$W/$\mu$m$^2$, for (top) 30 nm gold nanospheres, and (bottom) NHDs at identical linear absorption of $40$ m$^{-1}$. A shift is observed for the NHDs, indicative of self-propulsion. Solid lines are fits of the data using Eq.(2). \textbf{c)} Evolution of the normalized diffusion coefficient with the excitation intensity for NHDs (red circles) and 30 nm gold nanospheres (black squares) for identical solvent and absorption $\alpha=100$ m$^{-1}$. The dashed line is a guide for the eye, showing that the diffusion of isotropic gold nanospheres remains roughly constant. \textbf{d)} Estimated propulsion velocity for the NHDs as a function of the incident intensity, using Eq.(1) and constant $D(T)=D_0$ and $\tau_R$, determined from the value at equilibrium. The dashed line is a linear fit to the data, giving a slope of $3.8 \pm 0.4$ $\mu$m.s$^{-1}$/$\mu$W/$\mu$m$^2$.
    }
    \label{fig:Result2}
\end{figure}

From Eq.\eqref{eq:Deff}, we directly estimate the self-propulsion velocity $v$ [Fig.2d], which goes up to $\sim 150$ $\mu$m/s, \textit{i.e.}, $\sim 2 \cdot 10^3$ body-lengths per second, assuming constant $D^{NHD}_0$ and $\tau_R\propto R_h^2/D^{NHD}_0$. Importantly, this value is one order of magnitude faster than what phoretic microswimmers usually achieve \cite{Jiang_Sano-PRL-2010}, and already comparable to the highest velocities reported in the literature for nanoswimmers \cite{Lee_Fischer-NanoLetters-2014,Wang_vanHest-NatureComm-2024}.

 The high phoretic velocity originates from the strong asymmetric thermal gradient developed on the nanoscale silica lobe from the optical heating of the Au nanosphere. In the Rayleigh regime, the surface temperature elevation on the Au nanosphere is readily estimated as $\Delta T_S=I_{inc}\sigma_{abs}/4\pi\kappa R$, with $\sigma_{abs}$ the absorption cross section of the Au/SiO$_2$ NHDs \cite{Rings_Klaus-PRL-2010,Bregulla_Cichos-PRL-2016}, taken as $\sigma_{abs} = 1650$ nm$^2$ from full Mie computation, $\kappa=0.20$ W/m.K the thermal conductivity of the surrounding medium and $I_{inc}$ the incident intensity. From this, we compute $\Delta T_S \approx 2$ K at an intensity of $40$ $\mu$W/$\mu$m$^2$. Assuming that $v \sim \mu^{NHD} \Delta T_S/R_h$, a linear fit to the data gives a typical order of magnitude for $\mu^{NHD} \sim 1$ $\mu$m$^2$/s/K, in good agreement with values reported in the literature for the thermophoresis of silica spheres in aqueous solutions ($0.1\lesssim\mu\lesssim 10$ $\mu$m$^2$/s/K) \cite{Piazza-SoftMatter-2008,Braibanti_Piazza-PRL-2008,Pu_Benneker-SoftMatter-2023,Pu_Benneker-JCP-2024}. Further linking the translational active motion to the Stokes Einstein relation in an equilibrium system, we find that the (virtual) active temperature of the bath reaches $T^a = 6\pi\eta R_hD^{NHD}_\text{eff}/k_B\approx 600K$ at $\sim 40$ $\mu$W/$\mu$m$^2$, showing that the NHDs behave as fast nanomotors upon heating, and can already significantly overcome thermal noise.

\subsection{Self-propulsion in dense assemblies}
Having shown and quantified the individual propulsion of NHDs under low photothermal heating of the sample, we now turn to denser assemblies of nanomotors. We perform measurements over increasing concentrations. The results presented in [Fig.3a] for two sets of NHDs show a massive dependence of the effective diffusivity on their concentration, reaching a ratio $D^{NHD}_{\text{eff}}/D^{NHD}_0 \sim 12$ at the highest concentrations investigated ($c_0 \approx 4.7 \cdot 10^{14}$ particles/L). For each concentration, a monotonic dependence on the incident light intensity is observed, with a non linear behavior occurring as the concentration, incident intensity, and absorbed power increase.

We also compare our results to measurements on isotropic gold nanospheres at absorption matching that of the NHDs [Fig.3b]. While these nanospheres exhibit concentration-dependent diffusivity at a number density of particles above $\gtrsim 10^{14}$ NPs/L, the relative increase in $D^{Au}/D^{Au}_0$ remains lower than that observed for asymmetric NHDs at equivalent concentration. We ruled out convective effects as a potential bias on measurements, as confirmed by independent estimation of convective velocities in our sample, which remain typically under $\approx 100$ $\mu$m/s (see [Supplementary Sec.7, Supplementary Fig.12 \& 13] for details).
Instead, we now demonstrate that this strong increase, for isotropic Au nanospheres, results from a macroscopic change in the solvent temperature $\Delta \theta=T-T_0$, induced by the large-scale heating of the nanoparticles by the collimated, green Gaussian beams. This change in temperature arises from the slow decay $\propto 1/r$ of temperature from the surface of each light-to-heat nano-transducer. This effect results in a large temperature build-up upon increasing the concentration, an effect previously reported in 2D arrays of plasmonic nanoparticles but overlooked for thermophoretic nanomotors \cite{Baffou_Rigneault-ACSNano-2013}.

The change $\Delta \theta$ affects the viscosity $\eta(T)$ on a large scale and hence the Brownian diffusion coefficient (first term of r.h.s of Eq.\eqref{eq:Deff}). A key challenge is the correct estimation of $\Delta \theta$, which can be intuited at low power: considering the sample as an effective absorbing media, and neglecting convective heat transport, as suggested by the value of the Rayleigh number of the system, $\text{Ra} \lesssim1$ Supplementary Sec.7], we can reasonably assume $\Delta\theta$ as linear with the total absorbed power of the system, with:

\begin{equation}
    \Delta\theta=\gamma P_{abs} \approx \gamma \sigma_{\text{abs}} H c P_{\text{inc}},
    \label{eq:dtheta}
\end{equation} 
where $P_{abs}$ is the total absorbed power of the sample, $P_{inc}$ the incident power, $H$ the height of the solution chamber, and $\gamma$ an experimental factor accounting for the power-to-heat ratio, which we quantify in the following. Hence, the dependence of the data on $P_\text{abs}$ should translate to the dependence on the temperature.
From independent spectrometric measurements of the absorbance [Methods], we thus evaluate the total absorbed power for each sample. Subsequent plotting of the measurements of $D^{Au}/D^{Au}_0$ as a function of $P_{abs}$, for isotropic gold nanospheres, shows a universal collapse of all the data, independent of the incident power and concentration, with absorption coefficients ranging from $\alpha = 130 $ to $1385$ m$^{-1}$ [Fig.3b-top]. The collapse is also valid for Au/SiO$_2$ core-shell at different absorptions (see [Supplementary Fig.14]). This unambiguously shows that the dynamics of isotropic objects are dictated mainly by the \textit{macroscopic} temperature elevation. The sole effect of the microscopic surface temperature elevation $\Delta T_S$ remains negligible at the single particle level for isotropic particles, thus discarding any potential propulsion induced by the local modification of solvent properties or intrinsic shape asymmetry, as previously reported for gold nanoparticles in critical binary mixtures, for instance \cite{Rings_Klaus-PRL-2010,Schmidt_Volpe-NatureCommunications-2021}.

Importantly, the universal collapse of the data for isotropic nanoparticles with $P_{\text{abs}}$ provides a reference calibration to indirectly estimate the effect of global temperature elevation on the Brownian diffusion coefficient $D^{NHD}(T)$ of the NHDs, which notably evolves through the change of viscosity with the temperature, and affects the measured effective diffusivity $D^{NHD}_{eff}$ (see Eq.\eqref{eq:Deff}). From these results on isotropic Au nanospheres, we identify an empirical function $f(P_{\text{abs}})$:
\begin{equation}
    \frac{D^{Au}(T)}{D^{Au}_0} = \frac{D^{NHD}(T)}{D^{NHD}_0}=f[T(P_{abs})]\equiv f(P_{abs}).
\end{equation}

 The values of \AAu{$D^{Au}(T)/D^{Au}_0$ are well described with a second-order polynomial fit of the data [Fig.3b]. This is in agreement with a second-order development based on an Arrhenius law describing the viscosity dependence of binary mixtures on the temperature \cite{GRUNBERG_NISSAN-Nature-1949}[Methods]}.

 \begin{figure}[H]
    \centering
    \includegraphics[scale=1]{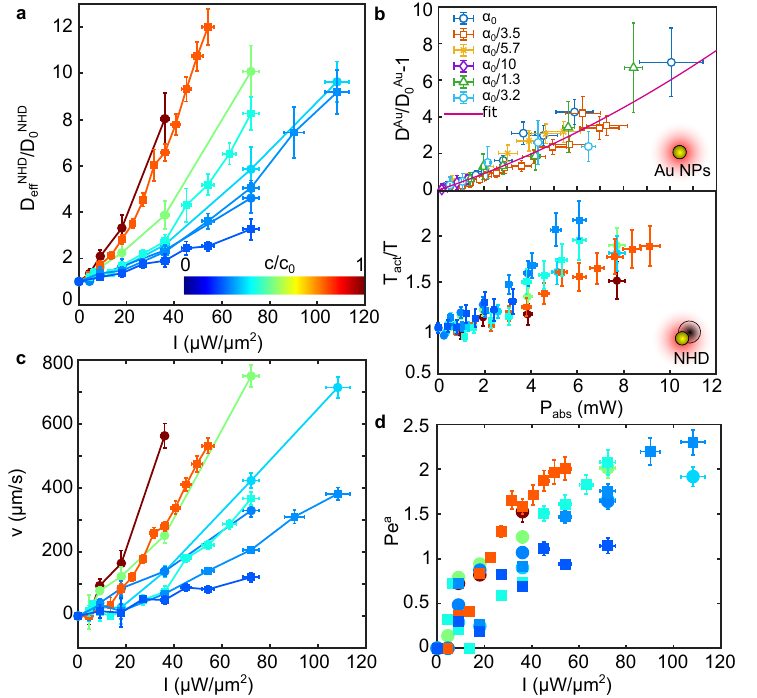}
    \caption{\textbf{Dynamics of nanomotors in dense assemblies.} \textbf{a)} Evolution of the diffusion coefficient of nanomotors at various concentrations, based on an initial concentration of NHDs $c_0 \approx 4.7\cdot 10^{14}$. particles/L$^{-1}$, corresponding to an absorption $\alpha= 770$ m$^{-1}$. \AAu{The data correspond to two batches of NHDs (circles and squares).} \textbf{b)} \textit{Top:} Evolution of the diffusion coefficient measured for 30 nm Au nanospheres for different incident power and initial absorption $\alpha$, with $\alpha_0=1385$ m$^{-1}$. All data collapse on the same curve when plotted against the total absorbed power $P_{abs}$. The data are captured by a second-order polynomial fit to the data, given by $f(P_{abs})-1$. \textit{Bottom:} Plot of the active temperature ratio to the fluid temperature $T^{a}/T=D^{NHD}_{eff}/D^{NHD}_0f(P_\text{abs})$ as a function of the absorbed power $P_{\text{abs}}$, evidencing a concentration dependency of the active dynamics of the NHDs. \textbf{c)} Estimation of the thermophoretic velocities $v$, using Eq.(5), from the data plotted in \textbf{a}. \textbf{d)} \AAu{Active P\'{e}clet number $ \text{Pe}^a=v.R_h/D^{NHD}_0f(P_\text{abs})$ computed for the different concentrations of the NHDs. At identical incident intensity, $\text{Pe}^a$ is higher for higher concentrations of NHDs.  \AAu{Color scales for \textbf{b-bottom}, \textbf{c}, and \textbf{d} are identical to \textbf{a}.}}
    \label{fig:Result3}}
\end{figure}

From Eq.\eqref{eq:Deff}, we can now account for the temperature dependence of $D^{NHD}(T)$ and $\tau^{NHD}_R(T)$. Data from NHDs show significant departures from isotropic objects, with a much higher increase in diffusivity. \AAu{This effect is evidenced when plotting the corresponding active temperature of the bath $k_BT^{a}/k_BT=D^{NHD}_{eff}/D^{NHD}(T)$, with $D^{NHD}(T)=D^{NHD}_0f(P_{abs})$, against $P_{\text{abs}}$ [Fig.3b-bottom]. Remarkably, the NHDs form an equivalent \textit{hot} bath, reaching up to two times the room temperature. The negative correlation between $T_{\text{act}}/T$ and the concentration is indicative of a specific effect of surface temperature elevation, itself modulated by the incident light intensity.}

Acknowledging again that $R_h = k_BT_0/6\pi \eta(T_0) D^{NHD}_0$, and noting that $D^{NHD}_{\text{eff}} \equiv D^{NHD}_{\text{eff}} (T,\Delta T_S)$, with $T=T_0+\Delta \theta=T(P_{abs})$ and $\Delta T_S=\Delta T_S(P_{inc})$, we estimate the thermophoretic speed of the nanomotors as:
\begin{equation}
    v[P_{\text{abs}}, P_{\text{inc}}] =\frac{18 \pi \eta_0}{k_BT_0}[D^{\text{NHD}}_0]^2f(P_{\text{abs}})\sqrt{\frac{1}{2} \left(\frac{T^a}{T}-1 \right)}
    \label{eq:v}
\end{equation}

The thermophoretic speed varies strongly with the concentration [Fig.3c], and increases monotonically. It reaches high values, up to $\sim 800$ $\mu$m/s, at the state-of-the-art of currently reported values for phoretic micro and nanomotors \cite{Lee_Fischer-NanoLetters-2014,Wang_vanHest-NatureComm-2024}. \AAu{The corresponding active P\'{e}clet number $\text{Pe}^a=v.R_h/[D^{NHD}_0f(P_\text{abs})]$ in [Fig.3d] shows that a higher activity compared to thermal diffusion can be obtained at high concentrations of NHDs, reaching values of up to $\text{Pe}\sim2.5$. Overall, our data are indicative of a temperature-dependent speed, regulated autonomously by the local concentration of the NHDs.}

\subsection{Thermal effects on self-propulsion}
Until now, we have only indirectly accounted for the macroscopic heating of the sample, based on the empirical knowledge of $f(P_{abs})$. We now rationalize our results from independent measurements and modeling of heating effects.
The dependence of the thermophoretic motion on temperature is a well-known effect \cite{Wurger-RPP-2010}, but its potential to control self-thermophoresis or the collective dynamics of nanomotors remains unexploited. Local changes in $T$ affect the solvent properties, the interactions of the fluid with the surface of a particle, and hence the value of its diffusiophoretic mobility $\mu$. 

\paragraph{Migration in a controlled temperature gradient.} The direct quantification of $\mu(T)$ from the thermophoretic speed of nanomotors is not straightforward, since we must account for the specificity of the nanoparticle shape to estimate $\mu(T)$ correctly from the velocity \cite{Jiang_Sano-PRL-2010}. We thus evaluate $\mu(T)$ and assess its temperature-dependence from independent measurements, by conducting thermophoresis experiments on SiO$_2$ microparticles in controlled temperature gradients
[Fig.4a]. To this end, we designed a thermophoretic cell composed of two metal sides, for which the temperature was set by two independent Peltier modules (Materials \& Methods). The experiments are performed in rectangular glass capillaries (height $50$ $\mu$m, length $500$ $\mu$m), with each metal side brought in contact with the edge of the capillary. It results in a constant horizontal temperature gradient across the channel width, typically set as $\Delta \Theta = 39$ K over $L= 500$ $\mu$m, leading to $\nabla T \approx 0.08$ K/$\mu$m, as quantified using Rhodamine B as a thermo-sensitive dye (Methods \& [Supplementary Fig.15]).

We follow the motion of SiO$_2$ microparticles and map their instantaneous velocity along the cell [Fig.4b,c]. The bare (electrostatically stabilized) SiO$_2$ microparticles are immersed in the same medium as for the NHDs. In this case, the microparticles migrate towards the hot side and reach maximum values of up to $\sim 1$ $\mu$m/s near the hot edge [Fig.4c]. However, determining the thermophoretic mobility of particles is not trivial, and must be carefully addressed to avoid measurement biases, particularly from flow advection, a generic issue to all phoretic phenomena \cite{Carrasco_Aubret-ACSNano-2025,Bregulla_Cichos-PRL-2016}. In particular, interfacial thermo-osmotic flows at the surface of the glass capillary significantly affect the motion of the particles and must be accounted for in the determination of the motion of silica beads. To this contribution, thermal convective flows could add up, induced by the thermal gradient across the cell of finite height. Ultimately, these two contributions can be combined in a single term $v_{\text{ad}}$. The final measured velocity is $v_{\text{SiO2}} = v_{\text{ph}} + v_{\text{ad}}$, where $v_{\text{ph}}=-\mu^{SiO_2}_T(T)\nabla T$ is the phoretic velocity to estimate.

Based on previous studies \cite{Bregulla_Cichos-JCP-2019,Bregulla_Cichos-PRL-2016}, we use gold nanoparticles to quantify $v_{ad}$. Thanks to their high conductivity, Au nanoparticles present a nearly homogeneous surface temperature, making them effectively insensitive to thermophoresis. As a result, with $v_{ph}\approx0$, they can act as probes for hydrodynamic flows.
We compared the motion of SiO$_2$ to $300$ nm gold particles across the capillary cell [Fig.4a-c]. In all the experiments, the Au nanoparticles migrated towards the hot side; however, their movement was slower than that of the SiO$_2$ microparticles, indicating a thermophilic behavior of silica under these experimental conditions. This behavior is in agreement with previous studies on silica microparticles, which, however, did not account for osmotic flows \cite{Pu_Benneker-SoftMatter-2023,Pu_Benneker-JCP-2024}. From statistical averaging of our data, we computed the thermophoretic mobility of silica using $\mu^{SiO_2}_T(T) \sim [v_{\text{SiO2}} -v_{\text{ad}}]/(\Delta \Theta/L)$ [Fig.4d].
The results present a strong variation of $\mu_T^{SiO_2}(T)$ with the temperature. We observe a nonlinear increase of $|\mu^{SiO_2}_T(T)|$ in the range $\Delta \theta = T-T_0=0-50$ K, up to a factor of 10. Importantly, the typical recorded values fall within the universal range reported in the literature for silica particles, with $0.1\lesssim |\mu^{SiO_2}_T(T)| \lesssim 10$ $\mu$m$^2$/s.K \cite{Wurger-RPP-2010}.

\paragraph{In situ probing of the temperature.} So far, we miss the direct link between $\Delta \theta$ and $P_{abs}$. To quantify this laser-induced temperature increase in solutions comprising NHDs, we exploit the temperature dependence of the Stokes-Einstein Brownian diffusion coefficient. Practically, we obtain $D^{SiO_2}(T)$ directly from the transverse fluctuations of the SiO$_2$ beads at different positions inside the capillary (Fig.4e \& Methods). The ratio $D^{SiO_2}(T)/D^{SiO_2}_0$ increases monotonically, as expected, reaching $\sim 4$ for $\Delta \theta= 40$K. 
We complemented our results with independent rheometric measurements of the evolution of the solvent viscosity with temperature ([Fig.4e] \& [Supplementary Fig.16]), which, despite experimental constraints limiting the temperature range, confirmed the trend and provided insights into solvent property modifications. Fitting our combined data \AAu{to an Arrhenius} equation for viscosity law \cite{GRUNBERG_NISSAN-Nature-1949}, expressed as $D^{SiO_2}(T)/D^{SiO_2}_0=\eta_0 T/\eta(T)T_0$ (see Material \& Methods), provides an analytical prediction for $D(T)/D_0$. From this knowledge, we can now measure the radial temperature profile inside a laser-heated solution of Au nanospheres, as shown in (see [Fig.4f] and [Supplementary Fig.17]). The temperature is obtained from the analysis of the diffusion of passive nanobeads by Fluorescence Correlation Spectroscopy (FCS), immersed in a solution of Au nanospheres, at different distances from the center of the laser spot. Linking the measurement of their diffusion to $\eta_0 T/\eta(T)T_0$, we extract the corresponding temperature. The result agrees with analytical calculations [Fig.4f,g], as described hereafter, and allows us to validate the procedure for extracting $\Delta \theta$.

We now have (i) a direct link between $\Delta \theta$ and $D(T)/D_0$ for passive Brownian objects [Fig.4e], and (ii) a direct link between $D(T)/D_0$ and $P_\text{abs}$ [Fig.3b], based on previous measurements on isotropic objects under laser heating. Hence, the sole knowledge of $P_\text{abs}$ allows us to directly evaluate the temperature inside a laser-heated solution of nanoparticles, irrespective of their size or shape. Assuming the recorded dynamics of isotropic objects is set solely by the macroscopic temperature $\Delta \theta$, we obtain a direct empirical evolution for $\Delta \theta(P_{abs})$ [Fig.4f-h]. Our experimental measurements indicate a typical \textit{macroscopic} temperature elevation of $\sim 8$ K for an absorbed power of $1$ mW in a $100$ $\mu$m cell height. The evolution of $\Delta \theta$ seems linear up to $P_\text{abs} = 5$ mW, then becomes sub-linear at higher powers [Fig.4f]. 

\begin{figure}[H]
    \centering
    \includegraphics[scale=0.9]{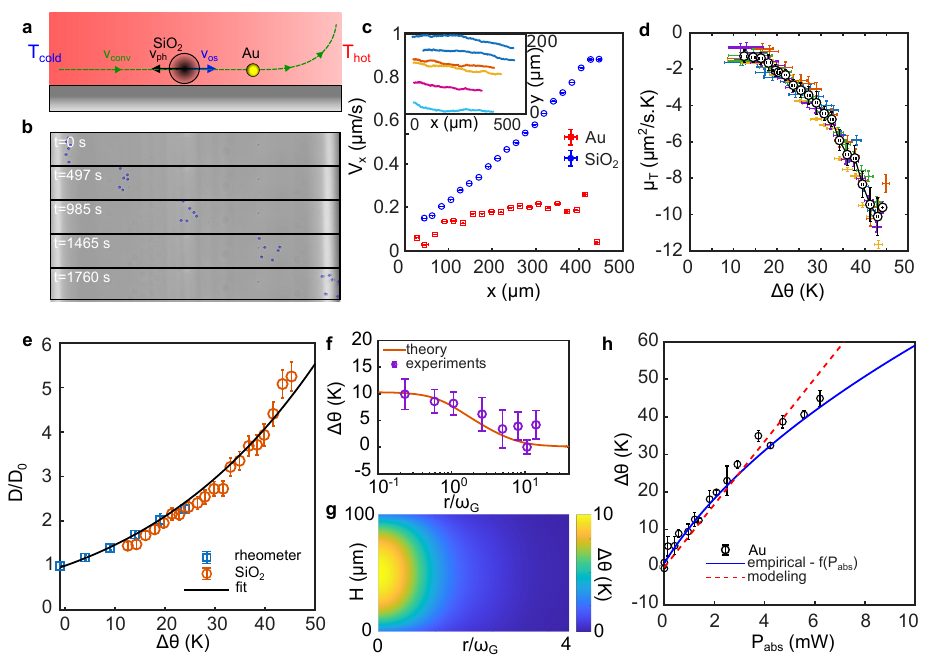}
    \caption{\textbf{Quantification of thermal effects from independent measurements.} \textbf{a)} Scheme of the experiment to measure the thermophoretic mobility of SiO$_2$ microbeads. A temperature gradient is applied horizontally across the cell, with $\Delta \Theta= T_{hot}-T_{cold} = 39$K. The contribution of osmotic and convective flows is accounted for by simultaneously tracking the motion of $300$ nm gold particles with $3.3$ $\mu$m microbeads. \textbf{b)} Snapshots of a typical experiment, showing the thermophilic behavior of SiO$_2$ colloids (in blue). \textbf{c)} Example of extracted velocities for the SiO$_2$ beads (blue circles) and 300 nm Au nanoparticles (red squares) across the channel width, and (\textit{Inset}) example of reconstructed trajectories of microbeads. \textbf{d)} Computed thermophoretic mobility $\mu_T^{SiO_2}(T)$ for 5 different experiments (dots), and corresponding averaged values (white circles), with $T=T_0+\Delta \theta$ and $T_0=296$K. \textbf{e)} Evolution of $D(T)/D_0$ for passive SiO$_2$ beads (red circles). Blue squares are calculated values from rheometric measurements of the viscosity of the solution. The solid line is a fit of combined data, using an Arrhenius law for the viscosity. \textbf{f)} Computed temperature profile (red line) and corresponding measurement at the center of the capillary, using fluorescent 210 nm fluorescent nanobeads and FCS to obtain their diffusion coefficient (blue circles). \textbf{g)} Analytical modeling: 2D plot of the temperature profile within the capillary, for $P_\text{abs}=1.2$ mW and $\omega_g=40$ $\mu$m. \textbf{h)} Temperature increase as a function of absorbed power, for $H=100$ $\mu$m and $\omega_g=40$ $\mu$m (blue circles), as extracted from the diffusion of Au nanospheres in a laser-heated solution (blue circles, binned average values of [Fig.3b]). The red dashed line shows the results obtained analytically from panel \textbf{g}. The solid blue line is an empirical estimation based on the parameters obtained from the Arrhenius law in panel \textbf{e}, and using $f(P_\text{abs}).$
    }
\end{figure}

\paragraph{Analytical calculations:} Our results are quantitatively validated by a simple analytical modeling of the laser heating of a solution of nanoparticles. We consider a minimal three-layers model, composed of successive glass-solvent-glass layers, and a Gaussian beam illumination with constant volumetric absorption ([Supplementary Fig.12 \& 13]). We solve the heat diffusion equation, assuming negligible convection (i.e. Rayleigh number $\text{Ra}\ll1$). A typical temperature profile is computed and presented in [Fig.4g] for $P_\text{abs}=1.2$ mW and $\omega_g=40$ $\mu$m. The thermal waist is found to be larger than the green optical waist by a factor of $\sim 5$, in agreement with previous studies \cite{Goy_Delabre-SoftMatter-2022}. Most of all, the theory establishes a first-order, linear relationship between $\Delta \theta$ and $P_\text{abs}$, with a slope of $\gamma =[8 \pm 0.5]$ K/mW, as defined in equation \eqref{eq:dtheta}, and in excellent agreement with the linear part of our experimental data [Fig.4h], achieved without any fitting parameters. At high powers, fluid advection may start to affect heat transport and should be considered in the heat diffusion equation for the solution.

\paragraph{Regulation of self-propulsion velocity}
\AAu{Eventually, we obtained measurements of the SiO$_2$ thermophoresis from both the nanometric self-propulsion [Fig.3c] and the migration of micrometric beads, with $\mu^{SiO_2}_T(T)$ [Fig.4d]. Our study of thermal effects now allows us to compare both sets of data as a function of the temperature. The above treatment of the data from SiO$_2$ microbeads implicitly assumed a linear relation of the phoretic velocity with the thermal gradient $v = -\mu_T^{SiO_2}(T) \nabla T$, with $\nabla T \sim 0.08$ K/$\mu$m being constant. For the NHDs, however, the thermal gradients are significantly larger, $\nabla T\sim \Delta T_s/R_h \approx 10 - 10^2$ K/$\mu$m and their variation with light intensity must be accounted for. Recent work indeed pinpointed the non-linearity of thermophoresis at high thermal gradients, as defined by the threshold condition $|R\mu(T)\nabla T|/D(T)>1$, and corresponding to the transition from a diffusive to a drift-dominated motion \cite{Mayer_Braun-PRL-2023}. Intriguingly, the self-propulsion regime of the nanomotors lies exactly around this critical value ($\text{Pe}^a \approx1$, see [Fig.3d]), and thus the linear regime may cease to be valid. To assess nonlinear effects, we compute an effective mobility of the nanomotors as $|\mu^{NHD}|=|v\cdot R_h/\Delta T_s|$. When plotted against the temperature, estimated from our previous analysis, data from microparticles and NHDs at varying concentrations exhibit different behavior [Supplementary Fig.18]. Therefore, $\mu^{NHD}$ depends not only on the temperature, but also on the local thermal gradient strength - a microscopic parameter, with $\mu^{NHD}(T,|\nabla T|)$. Combining all our datas from NHDs, we thus extract the evolution of $\mu^{NHD}(T)$ at fixed surface temperature elevations $\Delta T_S$. Remarkably, all data follow an identical trend with the temperature, similar to that of microscopic SiO$_2$ beads, with higher gradients yielding to lower mobilities [Fig.5a], as previously reported \cite{Mayer_Braun-PRL-2023}. Overall, this is indicative of a common physical origin, and enables, at first approximation, a decoupling of the nonlinear gradient dependence $\approx \Delta T_S/R_h$ from the ambient temperature dependence $T$. Then, we propose to write the effective thermophoretic mobility as a product of two independent terms $\mu^{NHD}(T,|\nabla T|) \approx \mu_T^{NHD}(T) \cdot \tilde{\mu}_{\nabla T}^{NHD}(\nabla T)$, with $\tilde{\mu}^{NHD}_{\nabla T}$ a dimensionless factor accounting for the non-linearity of thermophoretic propulsion.}

\AAu{Our previous data on SiO$_2$ microbeads now allow us to specifically extract the temperature dependent part $\mu^{NHD}_T$. From our measurements of $\eta(T)$, we obtain a negative Soret coefficient $S_T(T)=\mu_T^{SiO_2}(T)/D^{SiO_2}(T)$ [Fig.5a-inset]. Coincidentally, the data are well captured, in our range of accessible data, by a linear fit of $S_T=\alpha \Delta \theta$, at least as $\Delta \theta\gtrsim 10$ K, with $\alpha=-2.8 \pm 0.1$ K$^{-2}$ [Fig.5a] (see Methods). \AAub{We note that for identical surface and solvent interactions, and within the boundary layer approximation, the thermophoretic mobility does not depend on size \cite{Wurger-RPP-2010}. This picture is supported by the similar range of values for $\mu^{NHD}$ and $|\mu_t^{SiO_2}|$ observed in [Fig.5a]. Therefore, we extrapolate, from $S_T^{SiO_2}$, the Soret coefficient of the NHDs, $S_T^{NHD}=S_T^{SiO_2} D^{SiO_2}_0/D_0^{NHD}$}. Employing the linear response prediction, we can now test directly the relationship between $v$ and $\nabla T$. In particular, we link, from the definition of the above-mentioned active P\'{e}clet number (as shown in [Fig.3d]), and using $v=\chi |\mu^{NHD}|\Delta T_S/R_h$:}

\begin{equation}
    \text{Pe}^a= \frac{v \cdot R_h}{D^{NHD}(T)} = \chi \tilde{\mu}_{\nabla T}^{NHD} \cdot\frac{\mu_T^{NHD}}{D^{NHD}(T)} \cdot  \Delta T_S
\end{equation}
\AAu{where $\chi$ is introduced as a form factor, specific of the NHD geometry \cite{Jiang_Sano-PRL-2010}. }

\AAub{We then identify $\mu_T^{NHD}/D^{NHD}(T)= S_T^{NHD}(T)$, and plot the P\'{e}clet number as a function of the dimensionless number $S_T^{NHD}\Delta T_S\cdot = [\Phi(\Delta T_S)^2]/[\Phi_c(\Delta T_c)^2]$, which explicitly accounts for the volume fraction $\Phi$ and local gradient dependencies [Fig.5b]. We have defined the product $[\Phi_c(\Delta T_c)^2]={2R_h^3D_0^{NHD}/(3|\alpha| \gamma H\pi \omega_G^2 \kappa R^{Au}D_0^{SiO_2})} \approx 1.2\cdot 10^{-4}$ K$^2$ as a characteristic product of volume fraction and surface temperature elevation, specific to the system configuration, for which $\text{Pe}^a=\chi \tilde{\mu}^{NHD}_{\nabla T}$. Hence, for our reference concentration $c_0$ of NHDs, we get $\Delta T_c\approx 1.2$ K. As shown in [Fig.5b], all data from the NHDs show universal collapse, independent of $T$, $T_0$, $\Delta T_S$, $c$, or specific shape of the NHDs. The universal trend follows $\chi \tilde{\mu}_{\nabla T}$.}
\AAub{While noisy at low values of $\Phi\cdot(\Delta T_S)^2$, the data suggest a linear behavior [Fig.5b-inset] before becoming non-linear at a critical value $\Phi\cdot(\Delta T_S)^2 \sim \Phi_c\cdot(\Delta T_c)^2$. From the slope of the data in this linear regime, where $\tilde{\mu}_{\nabla T}\approx 1$ we estimate $\chi \approx 1.5 \pm 0.8$ for the form factor of the NHDs, of the same order of magnitude as previously reported value for microscopic Janus swimmers \cite{Jiang_Sano-PRL-2010}.}
\AAu{This remarkable result confirms the strong dependence of the thermophoretic velocity on absolute temperature and its local gradient, and the coupling between concentration and temperature that leads to a strong self-amplification of the speed. In particular, it demonstrates that higher volume fraction enables much lower laser  excitations than in the dilute regime to reach a comparable P\'{e}clet number.
Finally, the universal collapse observed in [Fig.5b] shows the robustness of our velocity measurements, and demonstrates, for the first time, that the 3D measurements of propulsion - enabled at the nanometric scale - allows one to generically assess the dependency of thermophoretic mobility on solvent, in the absence of nearby interfaces or individual tracking.}

\begin{figure}[H]
    \centering
    \includegraphics[scale=1]{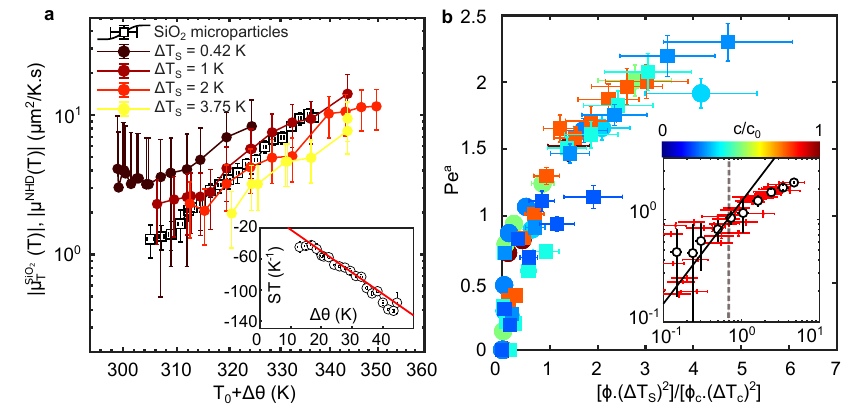}
    \caption{\AAu{\textbf{Regulation of self-propulsion with concentration}. \textbf{a)} Thermophoretic mobility of the SiO$_2$ microparticles, $\mu_T^{SiO_2}(T)$ (open black squares), and effective mobility of the NHDs, $\mu^{NHD}=|v.R_h/\Delta T_s|$ (circles). Data are plotted at fixed surface temperature elevations (\textit{i.e.}, $\Delta T_S$) from an average of raw data. Both microparticles and NHDs show similar trend with the temperature, with differing amplitude depending on $\Delta T_S$. \textit{Inset:} Extracted Soret coefficient $\mu_T(T)/D^{SiO_2}(T)$ of the microparticles, which shows a linear trend in the range of investigated temperature. Red solid line is a linear fit to the data, giving $y=\alpha\Delta \theta + \beta$, with $\alpha=-2.8 \pm 0.1$ K$^{-2}$ and $\beta=2 \pm 1$ K$^{-1}$. Hence, $\beta \ll \alpha \Delta \theta$ as $\Delta \theta\geq 10$ K. \textbf{b)} P\'{e}clet number of the NHDs, $\text{Pe}=v\cdot R_h/D^{NHD}(T)$ as a function of the dimensionless number $S_T^{NHD}\Delta T_s=[\Phi (\Delta T_S)^2]/[\Phi_c (\Delta T_c)^2]$, with $[\Phi_c (\Delta T_c)^2]$ a constant specific to the system geometry (see main text), giving $\Delta T_c=1.2$ K for a concentration $c_0$ of NHDs. All data show universal collapse on the same curve, for which the local slope is given by $\chi\cdot \tilde{\mu}_{\nabla T}$, representative of the nonlinearity of the thermophoretic velocity with the local gradient. Higher volume fraction $\Phi$ of the NHDs requires lower surface temperature elevation to reach equivalent P\'{e}clet, compared to small concentrations. \textit{Inset:} log-log plot of the main graph, suggesting the linear regime (black line) for small $[\Phi (\Delta T_S)^2]$ values with a slope giving $\chi \tilde{\mu}_{\nabla T}= [1.5 \pm 0.8]$. Black and white circles are binned average of all data (showed with red dots). Grey dashed line indicates the transition from linear to non-linear regime, identified for $\chi[\Phi (\Delta T_S)^2]/[\Phi_c (\Delta T_c)^2] \approx 1$, and therefore $\text{Pe}^a\approx 1$ (grey dotted line).}}
    \label{fig:Result5}
\end{figure}

\section{Discussion}
We have shown that Au-SiO$_2$ NHDs regulate their velocity in response to macroscopic heating of the solution, a process governed by the local nanoparticles concentration. This regulation leads to a positive feedback and self-amplification of their propulsion efficiency. \AAu{The overall dynamics of NHDs at a specific concentration can be rationalized by accounting for both the temperature dependence of the thermophoretic mobility and nonlinear effects, inherent to thermophoretic propulsion at the nanoscale.}
The resulting non-linear coupling between density and phoretic speed mirrors quorum sensing as observed in biology, a mechanism recently adapted in experiments with artificial microswimmers \cite{Baeuerle_Bechinger-NatComm-2018,Lefranc2025}. Unlike those systems, however, our findings demonstrate this behavior \textit{without} relying on external feedback for speed regulation. This discovery opens novel opportunities for actively tuning the density of 3D active baths, possibly leading to new exotic phases. 

Our P\'{e}clet analysis shows that activity can significantly dominate Brownian motion, especially in high-temperature samples, positioning thermophoretic dimers as good candidates for analyzing 3D emergent behavior. Previous theoretical studies have indeed suggested that thermophoretic dimers could exhibit dynamical aggregation, spontaneous density modulation, or swarming depending on the shape of the particles and sign of the interactions \cite{Wagner_Ripoll-EPJE-2021,Wagner_Ripoll-EPL-2017,RocaBonet_Ripoll-EPJE-2022}. While experimental confirmation of these phenomena is still pending, our study offers critical insights into the relevant parameters governing the collective dynamics of thermophoretic nanomotors. These results could further stimulate the development of refined theoretical models in this field.

Along this line, the self-regulation of velocity that we observe can potentially be tuned by the solvent and surface properties - a common feature of phoretic motion, which sets $\mu(T)$. So far, the thermophoresis of charged colloids in aqueous solution has been claimed to originate mainly from ionic and thermoelectric effects \cite{Wurger-RPP-2010,Pu_Benneker-SoftMatter-2023}, \AAu{in qualitative agreement with the negative Soret coefficient observed in our study.} Obtaining a complete physical picture of the molecular origin of self-thermophoresis is a complex problem. \AAu{As pinpointed in [Fig.5], a full description of thermophoretic nanopropulsion should account for non-local equilibrium, which is inherent to self-propulsion, together with the specificity of the solvent/particle interactions, as well as size and shape effects. Developing such a complete description remains a challenge, but could enable rational tuning of solvent properties, thus providing a new playground for modulating collective behavior at the nanoscale.}

To conclude, our findings reveal a remarkable adaptability in 3D assemblies of nanomotors, where local density can intricately tune the collective dynamics. Through collective heating, the temperature change retroactively acts back on the thermophoretic velocity of the particle. Beyond uncovering these fundamental interactions, our study paves the way for groundbreaking advancements in the dynamic control of active fluids, offering opportunities to engineer behavior in confined, three-dimensional environments. This work not only deepens our understanding of nanoscale collective phenomena but also unlocks innovative pathways for designing next-generation active materials and systems.

\color{black}
\newpage

\section{Methods}
\subsection{\textbf{Chemicals}}\label{sec:Chemicals}

30nm Gold Nanoparticles solution (GNPs, citrate buffer, ref. 741973, Sigma), 300 nm commercial Au nanoparticles (Sigma-Aldrich, 742082), 4-Mercaptobenzoic acid (4-MBA, $90\%$ technical grade), Polyacrylic acid (PAA, MW=250 K, $35$ wt\% in water) and Tetraethyl orthosilicate (TEOS, $<98$ \%) were obtained from Sigma-Aldrich. We used purified water (DI water, $18.2 M\Omega$.cm) for experiments. Ethanol (Et-OH, $>98$\%) and isopropanol (IPA, $>99.5$ \%, reagent grade) were purchased from Sigma-Merck. All glassware and magnetic stirring bars were cleaned with aqua regia (mixture of HCl and HNO3, [3:1]V) and then washed with DI water. All the chemical reagents were used as received without further purification.

\subsection{\textbf{Synthesis of nano-heterodimers}}\label{sec:JanSynth}
The heterogeneous nucleation of $SiO_2$ on $Au$ seeds is made from commercially available gold nanoparticles. The synthesis procedure is adapted from \cite{Park_Kane-Nanoscale-2018}. 1 mL of commercial 30nm-GNPs solution was centrifuged at 6000g for 20 min. The supernatant was removed to redisperse the nanoparticles in 1mL of DI water. In parallel, an aqueous solution of PAA (0.042 mM in DI water, 40 $\mu$L) and 4-MBA (5 mM in ethanol, 40 $\mu$L) was added to a mixture of 3.8mL IPA and 1.2mL MQ-water in a 10mL round-bottomed glass flask under vigorous stirring. After 10 minutes, the GNPs solution was added dropwise to the homogeneous medium, and kept under stirring for 30 minutes. 180 $\mu$L of aqueous ammonium hydroxide solution (28\%) was then injected, followed by 1.2 mL of TEOS solution (8.96mM in IPA) added dropwise. The reaction was kept under stiring for 12 hours. During the procedure, we kept the synthesis medium under stirring at a constant speed of 400 rpm. The final solution was centrifuged at 6000g for 10 mn to redisperse the products with a mixture of $[1:1]_{mol}$ IPA and DI water (respectively). 

\subsection{\textbf{Characterization of the NHDs}}
Typically, a few microliters of a solution ($\sim10$ $\mu$L) are deposited onto a TEM grid (CF300-Cu-50, Delta Microscopies) and then left to dry for at least 2 hours before analysis.
TEM images were then performed using a MET Jeol JEM-1400+ equipment from the Placamat platform (UMS 3626), operating at 120kV with a subnanometer resolution ($\sim0.4$ nm). 

\subsection{\textbf{Sample preparation}}
All samples of nano-heterodimers were prepared in glass capillaries of constant $100$ $\mu$m height (5012 - Rectangle Miniature Hollow Glass Tubing (VitroTubes™)), except for data in [Fig.2c \& 2d], collected in $H=200$ $\mu$m capillaries (3520, Rectangle Miniature Hollow Glass Tubing (VitroTubes™)).
All solutions are made of an isomolar cosolvent of 2-propanol and H$_2$O containing the initial solution of nanoparticles, giving a $pH \sim 9$. The solution is injected into the capillary, which is then sealed with capillary wax to prevent evaporation. Before the preparation, the NHDs (or gold and core-shell nanoparticles) are diluted at the required concentration, following the measurement of their absorbance with a spectrometer.

Experiments in gradient cells are performed in $H=50$ $\mu$m height capillaries (5005- Rectangle Miniature Hollow Glass Tubing (VitroTubes™)), with a width of $L=500$ $\mu$m to minimize thermal convection while warranting a strong enough temperature gradient. The solutions are made of commercial, bare (uncoated and electrostatically stabilized) SiO$_2$ microparticles of diameter $3.34$ $\mu$m (MicroParticles GmbH, SiO$_2$-R) re-dispersed in the same solvent as used for experiments with the nano-heterodimers. $300$ nm Au nanoparticles are introduced in the solution to evaluate the contribution of flow advection to the motion. Both the SiO$_2$ and Au particles were washed by centrifugation/redispersion cycles to remove the potential excess supernatant present in the solution before use. Typical final concentration give surface densities of the order of $\sim 100$ particles/mm$^2$ for SiO$_2$ and Au nanoparticles, necessary to enable individual tracking of each particle.

\subsection{Experimental methods}
\subsubsection{Confocal spectroscopy}
\paragraph{Experimental setup.} All experiments for the analysis of the dynamics of the NHDs were performed on a custom-made optical microscope. The microscope allows simultaneous heating using a green, CW collimated laser beam ($\lambda = 532$ nm, Laser 2000, CNI-DPSS-CW) and probing of the motion using a red, CW focused laser beam ($\lambda = 632$ nm, Thorlabs, HNLS008L-EC). The sample is mounted on a manual vertical translation stage, itself mounted on a manual XY translation stage (PI, M-545). The observation is made through a $\times 50$, long working distance, and infinity corrected objective (Olympus, NA=0.5), unless stated  otherwise, mounted on a piezoelectric scanner (Thorlabs, PFM450E) to enable accurate positioning in the sample of the cell. Alignment of the laser beam is observed directly on a CCD camera (Basler, V020A2A1920160UMB) using a 200 mm optical lens on the optical bath. The red probe laser is first enlarged using a telescope, and its polarization is rotated using a half-wave plate to allow its transmission through a polarizing beam splitter just before the observation objective. A quarter-wave plate is then used to obtain circularly polarized light and avoid potential polarization-dependent excitation. The red-laser is then focused through the objective, and we collect the back-scattered signal from the nanoparticles through the same objective. The light is polarization-filtered by the combination of the quarter-wave plate and the beam splitter, and focused in a single-mode optical fiber (Thorlabs, P1-460B-FC-2), playing the role of a pin-hole (mode-field diameter $\sim 4$ $\mu$m). The light is collected at the end of a fiber, spectrally filtered with a bandpass filter (Thorlabs, FLH633-5), before being sent to two single photon avalanche detectors (SPADs, Thorlabs -SPCM20A/M, PicoQuant - PD-050-CTE) with a 50/50 beamsplitter to avoid afterpulsing effects. The pulses are analyzed with a Time Correlated Single Photon Card (PicoHarp 300) with a temporal resolution of $4$ ps. We use the Time-Tagged-Time-Resolved (T3R) mode to reconstruct the cross-correlation function during post analysis, having access to all individual photon arrival times.
The heating of the nanoparticles is performed using the green laser. The fibered laser is collimated, and separated into two separate beams using a polarizing beam splitter and a half-wave plate. Both beams are focused at the back focal plane of identical, oppositely oriented $\times50$ Olympus objectives, giving collimated beams of waist $\omega_g = 40$ $\mu$m in the object plane and of identical intensity.

\paragraph{Data acquisitions}
All experiments are performed at the center of the capillary to avoid the strong reflection of the laser probe on the glass boundaries.
The correlation curves are computed directly from all the individual arrival times of photon events, using a home-made Matlab routine to extract T3R data from PicoQuant. We typically compute 10 points per time decade, extracting the correlation curves in a range from $10^{-8}$ to $1$ s. Typical acquisitions record $\approx 10$ mn of data, which can be extended in dilute samples where the signal-to-noise ratio is lower. For each $10$ mn time-trace, we extract at least $10$ correlation curves (one per minute), depending on the photon count, allowing us to obtain the dispersion of the extracted diffusion coefficients.

\subsubsection{Peltier module: experiments in controlled temperature gradients}
\paragraph{Experimental setup.} Experiments regarding the migration of $300$ nm gold nanoparticles and $3.3$ $\mu$m SiO$_2$ are performed on a commercial inverted microscope (Olympus IX73). The gradient cell is composed of two Peltier modules, on which we attached two metal plates of a contact section width of $\sim 700$ $\mu$m over the entire capillary thickness. A home-made PID controller allows for accurate setting of the temperature at the tip of the metal plates, with a typical accuracy of $0.1$ K. Peltier modules can be controlled independently from $-10$ \textdegree C to $+80$ \textdegree C. We use thermal grease to ensure continuous contact between the glass and the metal sides. All experiments were performed using a $\times20$ objective (Olympus, $NA=0.3$), enabling the observation of the entire capillary width and tracking of the motion of both SiO$_2$ microparticles and Au nanoparticles.

\paragraph{Data acquisition}
Data are recorded at 5 frames per second to enable the reconstruction of individual trajectories. A typical acquisition lasts between $20$ and $60$ mn, depending on the solvent composition. The temperature of each side of the Peltier is set at $T_c=20$ \textdegree C and $T_h=72$ \textdegree C, giving a $52$ K temperature difference between the two external edges. Direct measurement of the temperature profile using Rhodamine B yet indicates a gradient of $\Delta \Theta= 39$ K between $29$ and $68$\textdegree C on the capillary interior walls. For each acquisition, we perform a reversal of the gradient to check the reversibility of the motion, and to rule out sedimentation as a potential significant drift in our motion. All data are then analyzed using standard tracking routines under MATLAB. We compute the instantaneous velocities across the capillary for each particle. The velocity is computed as $v_i(\bar{x_i})=[x_{i+1}-x_{i}]/\Delta t$, where $\bar{x_i}=[x_{i+1}+x_{i}]/2$ and the subscript $i$ refers to frame number $i$. For each experiment, we bin all $v_i$ data to get $20$ data points along the channel width. We removed data located at $\pm20$ $\mu$m from the edge of the capillary, for which tracking was not accurate enough, due to the non-planarity of the surface. For each experiment, we then obtained the phoretic drift velocity of the silica microparticles as $ v_{ph}=v_{SiO_2}-v_{Au}$. For simplicity, we assume a linear temperature profile inside the chamber, a reasonable approximation as confirmed by the direct measurement of the temperature profile using Rhodamine B as a temperature probe [Supplementary Sec.9]. From this, the thermophoretic mobility coefficient is readily extracted as $\mu^{SiO_2}_T(T)= v_{ph}(x)L/\Delta \Theta$, with the temperature (in \textdegree C) at a position $x$ given by $T(x)\approx 29+x\Delta \Theta/L$. Error bars are obtained from the standard deviations $\sigma$ on the binned velocities such that $\delta v_{ph}=\sigma_{v_{SiO_2}}/\sqrt{N_{SiO_2}}+\sigma_{v_{Au}}/\sqrt{N_{Au}}$, with $N_{Au,SiO_2}$ representing the statistics obtained on each type of particle. The error on $\mu_T(T)$ is then given by $\delta\mu_T\approx \sqrt{[(L/\Delta T)\delta v_{ph}]^2+ [(v_{ph}/\Delta T)\delta L]^2+[(v_{ph}L/[\Delta T]^2)\delta \Delta T]^2}$. Similarly, the error on $T(x)$ is estimated as $\delta T(x) \approx \sqrt{[(1+x/L)\delta T_c]^2 + [(x/L)\delta T_h]^2 + [(\Delta T/L)\delta x]^2 + [(x\Delta T/L^2)\delta L]^2}$.

\subsection{\textbf{Correlation analysis from the time-traces}}
\paragraph{Expression of the cross-correlation function.}
The fitting of the cross-correlation curves is performed from the modeling of the electromagnetic field scattered by the nanoparticles and giving rise to intensity fluctuations on the SPADs. We adopt to this end a theoretical approach initially developed for analyzing light scattering in gels, which accounts for a possible constant reference electromagnetic field due to static inhomogeneities, similar to heterodyne measurements \cite{Puset_vanMegen-PhysicaA-1989,Joosten_Pusey-Macromolecules-1991}. Briefly, the intensity autocorrelation function is given by:
\begin{equation}
    g^{(2)}(\tau)=\frac{\langle I_1(t+\tau)I_2(t)\rangle}{\langle I_1(t)\rangle \langle I_2(t)\rangle}
\end{equation} with $I_1(t)$ and $I_2(t)$ the intensities at time $t$ for the SPAD 1 and 2 (respectively). \\
Writing the total field as the sum of a constant field $E_c(t)$ and the field scattered by the nanoparticles $E_s(t)$, one can develop the full expression for $g^{(2)}$ (see [Supplementary Sec.3] for a more detailed derivation).
Our confocal configuration gives rise to two distinct terms in the autocorrelation curve.\\
The first one results from the coherent fluctuations of interfering waves from multiple particles, for which the distance between scatterers changes in time. The characteristic timescale of fluctuation is given by $\tau_C=1/2Dq^2$, where $q=(4n\pi/\lambda)\sin{\theta/2}$ is the scattering wavevector. In our case, $n\approx 1.37$ as measured by refractometer, $\lambda =632.8$ nm, and $\theta=\pi$ for back-scattered photons. This component corresponds to the traditional exponential decay used in the Dynamic Light Scattering (DLS) experiment for normally diffusing particles:

\begin{equation}
    g^{(2)}_C(\tau)=\beta \exp{\bigg(-\frac{\tau}{\tau_C}\bigg)}
\end{equation} where $\beta \lesssim 1$ is an experimental factor and $\tau_C$ is the characteristic time for a particle with a diffusion coefficient $D$ to travel a distance $1/q$. 

The second, incoherent term is related to particle number fluctuations in the confocal volume. The contribution to $g^{(2)}$ corresponds to the traditional one used in Fluorescence Correlation Spectroscopy and is given, in the absence of any background field or noise, by: 
\begin{equation}
    g^{(2)}_I(\tau)=G_0\bigg(1+\frac{\tau}{\tau_D}\bigg)^{-1}\bigg(1+k^2\frac{\tau}{\tau_D}\bigg)^{-1/2}
\end{equation} where $G_0$ is an experimental parameter that scale as $1/N$, $\tau_D=\omega^2/4D$ is the characteristic time spent by a particle with a diffusion coefficient $D$ in the confocal volume of size $\omega$ and $k=\omega_{z}/\omega$ the confocal radius ratio in the z direction and the x-y plan ($\omega_{z}$ and $\omega$ respectively).

The total autocorrelation curve is given by:
\begin{equation}
    g^{(2)}(\tau)-1=g^{(2)}_{C}(\tau)+g^{(2)}_{I}(\tau)
\end{equation}

In effect, static reflections on the glass interfaces give rise to a static field contribution, as mentioned above. This yields a first-order correction to the coherent time as \cite{Joosten_Pusey-Macromolecules-1991} $(\tau_C')^{-1}=(\tau_C)^{-1}[1-\sqrt{1-\beta}]/\beta$ (see [Supplementary Sec.3]).

\paragraph{Fitting procedure.} For each experimental curve, we determine the plateau of the cross-correlation function by averaging all data points from $\tau=10^{-8}$ to $10^{-6}$ s. The obtained value gives the total amplitude $A=\beta+G_0$. The waists $\omega$ and $\omega_Z$ of the laser are determined from independent measurements, giving $\omega=550$ nm, and $\omega_Z=19\omega$ [Supplementary Sec.4]. It results in only two fitting parameters $\{D,G_0\}$, independent of the concentration of nanoparticles in the sample.

\subsection{Calculation of the self-propulsive velocity}
The dynamics of self-propelled particles is quantified from their effective diffusion $D_{eff}=D(T)+v^2\tau_R/6$, with $D(T)$ the equilibrium diffusion coefficient measured at temperature $T$. This is justified by the time and space scales probed in our experiments, occurring at low P\'{eclet numbers}, unable to resolve the persistence length of the particles \cite{Kurzthaler_Franosch-SciRep-2016}. Our analysis relies on the fact that $D(T)/D_0=f(T)$ is independent of the type of nanoparticles used. The corresponding ratio for NHDs can therefore be determined independently from the ratio obtained with isotropic nanoparticles. We assume that the reorientation time $\tau_R$ is independent of the activity, with $\tau_R(T)=4R_h^2/3D(T)$. Furthermore, the hydrodynamic radius is estimated and measured directly from the equilibrium diffusion coefficient at room temperature, with $R_h=k_BT_0/6\pi\eta_0D^{NHD}_0$, and $\eta_0=2.45 $ mPa.s determined independently from rheometric measurements.
Rearranging the expression of $D_{eff}$, we readily obtain Eq.\eqref{eq:v}.
The error $\delta v$ on $v$ is obtained directly from eq.\eqref{eq:v} as:
\begin{equation}
\delta v=\sqrt{\sum_i^N \left[\left |\pdfrac{v}{x_i}\right|\delta x_i\right]^2}
\end{equation}
where the $\{x_i\}$ are all variables required for the computation of $v$, namely $\{\eta_0,T_0, D_0^N, D_{eff}^N\}$

\subsection{Extraction of $P_{abs}$}
In our experiments, $f(T)\equiv f(P_{abs})$ is extracted using a second order polynomial fit. The fit is performed by weighting each point with its error bar. This type of fit is in agreement with a first-order development based on the Arrhenius law for the viscosity \cite{GRUNBERG_NISSAN-Nature-1949}. Taking the effective viscosity of the mixture as $\ln \eta \approx N_1 \ln \eta_1 + N_2 \ln \eta_2 + N_1N_2c$, with $c$ a constant, $N_i$ the mole fraction of constituent $i$, and $\eta_i$ the corresponding viscosity, we can develop $T/\eta (T)$ for small values of $\Delta \theta=T-T_0$. In our case, $N_1=N_2=N$. Assuming the $\eta_i$ follow a single component Arrhenius law, one immediately gets:
\begin{equation}
    \frac{T\eta_0}{\eta(T)T_0}\approx{1+\alpha' \Delta \theta + \beta' \Delta \theta^2+\cdots}
\end{equation}

 From the relation $\Delta \theta = \gamma P_{abs}$, we find $f(P_{abs})=1+a P_{abs}+bP_{abs}^2$, with $a=0.42 \pm 0.05$ mW$^{-1}$ and $b=0.02 \pm 0.01$ mW$^{-2}$, and the uncertainty extracted from bootstrap statistics. This provides us with an analytical (empirical) function to quantify $v$ in dense assemblies.

\subsection{Calculation of the effective mobility and Soret coefficient}
\AAu{\paragraph{Effective mobility $\mu^{NHD}$.} Data from [Fig.5a] are obtained by performing a running average over all data from NHDs falling within a specific interval of values for $\Delta T_S=\sigma_{abs}I_{inc}/4\pi\kappa R^{Au}$, with $R^{Au}=16 \pm 2$ nm, $\kappa= 0.2$ W/m.K, and $\sigma_{abs}=1650 \pm 100$ nm$^2$. We selected the intervals $\Delta T_S=0.42 \pm 0.32$ K, $\Delta T_S=1.12 \pm 0.38$ K, $\Delta T_S=2.0 \pm 0.5$ K, and $\Delta T_S=3.75 \pm 1.25$ K. All data for a specific $\Delta T_S$ value are then filtered using a uniform filter to smooth the data, with a window of $5$ K. Error bars on the data are obtained from the same process applied to the initial set of values $\mu^{NHD}\pm\delta \mu^{NHD}$.}
\AAu{\paragraph{Soret coefficient.}
The Soret coefficient of SiO$_2$ microbeads is computed directly by dividing the measured $\mu_T^{SiO_2}$ by the computed value of $D^{SiO_2}(T)$ using the Arrhenius law. Fit to the data gives $S_T=\alpha\Delta \Theta +\beta$, with $\alpha=-2.8 \pm 0.1$ K$^{-2}$ and $\beta=2 \pm 1$ K$^{-1}$. In most of our range of accessible values for $\Delta \theta \gtrsim 10$ K, we thus have $\alpha\Delta \Theta \gg \beta$, and we set $\beta\approx 0$.}

\paragraph{P\'{e}clet calculation.} The dimensionless number $S_T^{NHD}\Delta T_s$ is then given by $\alpha\Delta \theta (D_0^{SiO_2}/D_0^{NHD})$, with, from Eq.(3), $\Delta \theta=\gamma \sigma_{abs}c H I\pi \omega_G^2/2$. Expressing $\sigma_{abs}$ as a function of $I$ and $\Delta T_S$, we find $S_T^{NHD}\Delta T_s=[\Phi (\Delta T_S)^2]/[\Phi_c (\Delta T_c)^2]$. \\
We identify $[\Phi_c (\Delta T_c)^2]=2R_h^3D_0^{NHD}/(3|\alpha| \gamma H\pi \omega_G^2 \kappa R^{Au}D_0^{SiO_2})$ as a characteristic value for the product of volume fraction and surface temperature elevation, specific of the NHD and system geometry. Typically, $\text{Pe}^a\approx 1 $ when $[\Phi (\Delta T_S)^2]=[\Phi_c (\Delta T_c)^2]/\chi$.

\subsection{Measurement of the transverse diffusion coefficient}
The transverse diffusion coefficient of SiO$_2$ in thermophoretic cells is given by computing the second moment of the bead displacement along the vertical $y$ direction on the image. We calculate $<[\Delta y_j]^2>=<[y_i-y_{i+j}]^2>_i$ for $j\in[1;10]$, where the subscripts refer to frame numbers, recorded at $5$ fps. All data $[\Delta y_j]^2/\sqrt{\tau_j}$ collapse on the same graph, as expected for Brownian motion, which allows us to directly extract the Brownian diffusion coefficient at various positions across the capillary. Further linking $x$ to $\Delta \theta$, the temperature dependency of Brownian diffusion in the isopropanol-water cosolvent is readily extracted (see [Fig.4e], main text).

\subsection{\textbf{Rheometric measurements}}
\paragraph{Experimental setup} The direct measurement of the viscosity of the 2-propanol/water mixture is performed on a commercial rheometer (HR20 from TA Instruments\copyright). By using a cone-plate geometry, we measured the shear stress $\sigma_\text{shear}$ at different imposed shear rate $\dot\gamma_\text{shear}$ to deduced the viscosity by the linear relation $\sigma_\text{shear}=\eta\ \dot\gamma_\text{shear}$. The temperature control module allows the setting of the temperature up to $\sim 45$ \textdegree C, above which evaporation cannot warrant the reliability of the measurements.

\paragraph{Data acquisition and analysis} For each fixed temperature, we performed measurements for $\dot\gamma_\text{shear}$ between 50 and 300 s$^\text{-1}$ with a typical acquisition time of $60$ s on a small volume of 2-propanol/water mixture ($\sim200$ $\mu$L). We then obtained a distribution of viscosity for $T=20-45^\circ$C, which allowed us to model an empirical law as a function of temperature. 
The cumulated data from the diffusion of micrometric silica beads and from rheometry [Fig.4d] are fitted based on an Arrhenius law for viscosity. From the viscosity values measured by rheometry, we compute $T \eta_0/T_0 \eta(T)$. Combined data are fitted with the function $F(T)= T\eta_0/(T_0 \eta_\infty \exp[{B/T}])$, using $B$ as the only fitting parameter, since $\eta_\infty$ is fixed from the condition $D(T_0)/D_0=1$. The fit of the data gives $B=3145$ K, for $\eta_0=2.45$ mPa.s and $T_0=294.15$ K.

\subsection{Measurement of the optical and thermal properties of the solution}
\paragraph{Absorbance} Absorption spectra are measured using an absorbance spectrometer in either $100$ $\mu$m wide quartz cuvette or $1$ cm plastic cuvettes. The measurements are acquired in the range $400$ nm - $800$ nm. The absorbance is extracted from the transmission values at $\lambda =532$ nm, for both isotropic nanoparticles and NHDs, as given by $A=\alpha H/\ln(10)$, with $\alpha=c \sigma_{\text{abs}}$ the linear absorption. Knowing $H$, and through the estimated value of $\sigma_{abs}=1650$ nm$^2$ for a 30 nm Au nanosphere, we can thus infer the typical concentration of nanoparticles. Both solutions of nano-heterodimers and Au nanoparticles show a plasmon resonance wavelength around $530$ nm, with a small redshift for the $Au-SiO_2$ solution, as expected by the presence of a higher refractive index medium around gold nanospheres.

\paragraph{Refractometry}The refractive index of the solvent is measured using an Abbe refractometer available in the lab, giving $n=1.37$. 

\paragraph{Thermal conductivity}
For an binary mixture made of two solvents, the thermal conductivity can be predicted from the knowledge of the mass fractions as $\kappa=m_i\kappa_i+m_j\kappa_j$, with $\kappa_i$ and $\kappa_j$ the thermal conductivities, and $m_i$ and $m_j$ their respective mass fraction. We compute $\kappa=0.24$ W/K/m, in good agreement with previous experimental studies, which reported values around $\kappa\simeq0.20$ W/K/m, with a non linear evolution of $\kappa$ with the mass fractions \cite{Assael_Wakeham-IJT-1989,Iervolino_Sarro-ThermochimicaActa-2009}. We use the value from the literature for the analytical calculations performed in this study.


\newpage
\section*{Acknowledgments}
The authors thank M. Perrin, T. Gu\'{e}rin, and C. Pin for fruitful discussions. We also thank Y. Louyer for lending the PicoHarp, and T . Salez (EMetBrown Lab) for the rheometer.\\

This research was funded, in whole or in part, by l'Agence Nationale de la Recherche (ANR), project NanoDArt, ANR-24-CE30-6855-01. For the purpose of open access, the author has applied a CC BY public copyright licence to any Author Accepted Manuscript (AAM) version arising from this submission.
This project has received funding from the European Union’s Horizon 2020 research and innovation programme under the Marie Skłodowska Curie grant agreement No 886024. TEM observations were performed at Plateforme Aquitaine de Caractérisation des Matériaux(PLACAMAT, CNRS-UAR 3626). 

\section*{Authors contributions}
AA conceived the project. AA and YdF, and SC developed the experimental setup. YdF performed the experiments. YdF, MR, and MHD performed the synthesis of the nanomaterials. YdF, AA, UD, and JPD modeled and analyzed the data. AA and YdF wrote the manuscript. All authors reviewed the manuscript and contributed to the research work presented in the manuscript.

\section*{Competing Interests}
The authors declare no competing interests

\end{document}